\documentclass[letterpaper,11pt]{article}

\PassOptionsToPackage{nosort}{cite}

\usepackage{amssymb}
\usepackage{amsmath}
\usepackage{trimclip}
\usepackage{mathtools}
\usepackage{dsfont}
\usepackage{mathrsfs}
\usepackage{enumitem}
\usepackage[justification=centering]{caption}
\usepackage{float}
\usepackage{tikz}
\usepackage{verbatim}
\usepackage{multirow}
\usepackage{youngtab}
\usepackage[T1]{fontenc}
\usepackage{chet}

\allowdisplaybreaks

\definecolor{darkblue}{rgb}{0.1,0.1,0.7}
\hypersetup{colorlinks,
           linkcolor={darkblue},
           citecolor={darkblue},
           urlcolor={darkblue},
           breaklinks=true,
           pdftitle={Fine Spectrum from Crude Analytic Bootstrap},
           pdfdisplaydoctitle
}

\newcommand{\lsp}{\hspace{0.5pt}}
\newcommand{\lnsp}{\hspace{-0.5pt}}

\renewcommand{\geq}{\geqslant}

\def\cG{{\mathcal{G}}}

\def\cM{{\mathcal{M}}}

\def\cR{{\mathcal{R}}}
\def\cT{{\mathcal{T}}}

\def\cV{{\mathcal{V}}}

\DeclareMathOperator{\tr}{tr}

\newcommand{\bea}{\begin{eqnarray}}
\newcommand{\eea}{\end{eqnarray}}
\newcommand{\beq}{\begin{equation}}
\newcommand{\eeq}{\end{equation}}
\newcommand{\bal}{\begin{equation}\begin{aligned}{}}
\newcommand{\eal}{\end{aligned} \end{equation}}

\preprint{DESY-25-039}

\title{Fine Spectrum from Crude Analytic Bootstrap}
\author{Jake Belton,$^{\!a}$\footnote{\href{mailto:jake.belton@kcl.ac.uk}{jake.belton@kcl.ac.uk}} 
Nadav Drukker,$^{\!a}$\footnote{\href{mailto:nadav.drukker@gmail.com}{nadav.drukker@gmail.com}} 
Ziwen Kong,$^{\!b}$\footnote{\href{mailto:zwn.kong@gmail.com}{zwn.kong@gmail.com}}
and Andreas Stergiou$^{a}$\footnote{\href{mailto:andreas.stergiou@kcl.ac.uk}{andreas.stergiou@kcl.ac.uk}}
}

\affiliation{$^a$Department of Mathematics, King's College London, Strand, London, WC2R 2LS, United Kingdom\\[2mm]
$^b$Deutsches Elektronen-Synchrotron DESY, Notkestr.\ 85, 22607 Hamburg, Germany}

\abstract{The magnetic line defect in the $O(N)$ model gives rise to a non-trivial one-dimensional defect conformal field theory of theoretical and experimental value. This model is considered here in $d=4-\varepsilon$ and the full spectrum of defect operators with dimensions close to one, two and three at order $\varepsilon$ is presented. The spectrum of several classes of operators of dimension close to four and operators of large charge are also discussed. Analytic bootstrap techniques are used extensively, and efficient tools to deal with the unmixing of nearly degenerate operators are developed. Integral identities are also incorporated, and it is shown that they lead to constraints on some three-point function coefficients and anomalous dimensions to order $\varepsilon^2$.}

\date{March 2025}

\begin{document}

\maketitle

\toc

\section{Introduction}
\label{sec:intro}
Conformal defects provide a rich framework for studying conformal field theory (CFT), introducing localised perturbations to a bulk CFT while preserving a subset of the original conformal symmetry. Their theoretical interest is supplemented by their extensive applications in physical systems with impurities, boundaries, or interfaces, making them relevant for condensed matter and statistical physics. Localised excitations on the defect, known as defect operators, obey much of the same properties of operators in CFT, including the operator product expansion (OPE) and crossing symmetry of correlation functions. This structure of the defect CFT (dCFT) enriches the scope of the theory, provides a natural setting for applying and extending bootstrap techniques and enables new insights into how CFTs behave under deformations.

A particularly important example of dCFT, on which we focus this work, is the line defect of the $O(N)$ model, which was recently discussed in \cite{Cuomo:2021kfm} following older work \cite{Hanke:2000, Allais:2014fqa}.  This is a one-dimensional dCFT, that arises in the infrared of an RG flow that is triggered by a line deformation of the $O(N)$ CFT. More specifically, if $S_{O(N)}[\phi]$ is the action of the $O(N)$ model in $d=4-\varepsilon$, the defect deformation is of the form
\begin{equation}\label{eq:SON_deform}
    S_{O(N)}[\phi]\to S_{O(N)}[\phi]+h_0\int_{-\infty}^\infty d\tau\,\phi_N(\tau,\mathbf{0})\,,
\end{equation}
where we integrate over one dimension only, $h$ is the defect coupling and $\phi_N$ the $N$-th scalar field in the $\phi$ $O(N)$ multiplet. With the bulk theory remaining at its fixed point, a renormalisation group flow accessible with perturbative techniques in $d=4-\varepsilon$ for $\varepsilon$ small ensues, for the deformation in \eqref{eq:SON_deform} is relevant. The order $\varepsilon$ beta function of the coupling $h$ can be easily computed to be \cite{Allais:2014fqa, Cuomo:2021kfm}
\begin{equation}\label{eq:beta_h}
    \beta_h=-\tfrac12\varepsilon h\big(1-\tfrac{1}{N+8}h^2\big)\,,
\end{equation}
with non-trivial zero at $h^2_*=N+8+\text{O}(\varepsilon)$. 
A variety of results for this theory were obtained in \cite{Cuomo:2021kfm} and \cite{Gimenez-Grau:2022czc}. Significantly expanding on that, we provide an extensive account of the low-lying spectrum of operators in this theory.

Our calculations of the perturbative spectrum of the $O(N)$ magnetic line dCFT in $d=4-\varepsilon$ leverages Feynman diagrammatic and analytic bootstrap tools. The spectrum of operators of dimension close to one and two was computed in \cite{Gimenez-Grau:2022czc} using Feynman diagrams. There is no obstacle to continuing in this vein to higher-dimension operators, but we take a slightly different approach. We use Feynman diagrams to evaluate four-point functions of all operators up to dimension close to two, and then develop an efficient perturbative bootstrap algorithm to extract results for exchanged operators up to dimension close to four using analytic bootstrap. We find the full spectrum at dimension close to three and several subclasses of operators of dimension close to four.

The conformal bootstrap relies on a variety of tools to study CFTs. Its numerical implementation, pioneered in \cite{Rattazzi:2008pe}, shows its power in addressing strongly coupled theories. Analytic techniques based on the bootstrap principles have a variety of applications also at weakly-coupled settings. In particular, analytic bootstrap techniques have been extensively applied to the $\varepsilon$ and large-$N$ expansions of the $O(N)$ model \cite{Alday:2017zzv, Henriksson:2018myn, Alday:2019clp, Bertucci:2022ptt}, as well as other models \cite{Henriksson:2020fqi, Henriksson:2021lwn, Kousvos:2022ewl}.

The analytic bootstrap approach was eventually systematised using the inversion formula \cite{Caron-Huot:2017vep}. One would ideally seek to employ such dispersion relations and inversion formulae to study the spectrum of dCFTs, as these tools have proven to be quite efficient and powerful for CFTs in $d\geq 2$. While significant effort has been invested in understanding dispersion relations for CFTs in $d=1$ \cite{Mazac:2018qmi, Paulos:2020zxx, Bonomi:2024lky, Carmi:2024tmp}, the kernels that feature in them have not been computed in general. Faced with a lack of sophisticated methods applicable to our case, then, we resort to the cruder approach mentioned above that nevertheless proves to be quite powerful.

A famous problem one encounters in the analytic bootstrap is how to distinguish operators that are nearly degenerate, known as the \emph{unmixing problem}. This is similar to the problem of degenerate perturbation theory in quantum mechanics, except that in this case we may not know a priori what the dimension of the space of states is. In the case of the Wilson line of $\mathcal{N}=4$ super Yang--Mills, the unmixing problem was recently discussed \cite{Ferrero:2023znz, Ferrero:2023gnu}. 

If one focuses on a single four-point function, it is often possible to reproduce it by assuming the exchange of one operator near every even or integer dimension. The dimension and structure constant one finds this way are in fact weighted averages of the real spectrum. Requiring that multiple four-point functions exchange the same set of intermediate states gives a set of relations on the real spectrum, via its multiple averages. This problem gives rise to a set of polynomial equations in multiple variables, for which in general there are no efficient solution techniques. One of our main advances is to recast these equations in matrix form, making the solution straightforward. This representation also makes it evident how many operators mix, so we can be sure that we have fully unmixed them.

A further tool that we use is the integral identity of~\cite{Drukker:2022pxk} and generalisations thereof. The idea behind the identity presented there relies on the fact that a symmetry breaking defect, like the magnetic line breaking $O(N)\to O(N-1)$, gives rise to a space of defects, or defect conformal manifold, which in this case is $S^{N-1}=O(N)/O(N-1)$. The related defect marginal operators are known as tilts, $t_a$, and one can represent the curvature of $S^{N-1}$ in terms of an integral of their connected four-point function
\bal\label{tilt-id}
    R_{abcd}=2 \int d\chi\log|\chi|& \langle t_a (1) t_b(\chi)t_c(0)t_d(\infty)\rangle_c
    =(\delta_{ac}\delta_{be}-\delta_{ae}\delta_{bc})\langle t_e(0)t_d(\infty) \rangle\,.
\eal
The second equality uses that $S^{N-1}$ is a symmetric space and this identity is a non-trivial constraint, since the curvature of the sphere 
is completely determined in terms of its radius.

This identity was applied to the $O(N)$ model in \cite{Sakkas:2024dvm} and here we use a generalisation of it for operators other than the tilt. Operators carrying $O(N-1)$ charge $J$ can be viewed as part of a line bundle on the conformal manifold, specifically in the $J$'s tensor product of the tangent bundle, see~\cite{Baggio:2014ioa,Baggio:2017aww}. In terms of \eqref{tilt-id}, replacing the operators at 0 and $\infty$ with charge $J$ operators, the identity still holds with a factor of $J$ on the right-hand side due to the number of allowed index contractions as in \eqref{integral-id}.

These identities have a variety of applications, but in this paper we use them as a tool to study the spectrum and structure constants of defect operators. As observed in \cite{Cavaglia:2022qpg}, the naive application of such identities in perturbation theory fails. As explained in~\cite{4-pot}, this is due to unambiguous contact terms in correlation functions, which are absent from the four-point function at separated points. These contact terms are given by ratios of perturbative structure constants to anomalous dimensions, hence contributing to the theme of this work.

We begin with a quick review of the model in the next section, where most of our notations and conventions are also defined. Our analytic bootstrap methods are described in Section~\ref{sec:analytic}, where our approach to the unmixing problem is also developed. The spectrum we obtain for our theory is presented in Section~\ref{sec:spectrum}. The application of the integral identities to the large charge sector is explored in Section~\ref{sec:integrals}. We conclude in Section~\ref{sec:conclusion}.

\section{The magnetic line dCFT}\label{sec:review}
The magnetic line dCFT in $d=4-\varepsilon$ dimensions is described by the infrared fixed point of the action
\begin{equation}\label{eq:action} 
    S=\int d^{\lsp 4-\varepsilon}x\, \big( \tfrac{1}{2}\partial^\mu \phi_a\partial_\mu\phi_a+\tfrac{1}{2}(\partial_\mu \phi_N)^2 +\tfrac{1}{4!}\lambda_0(\phi_a^2+\phi_N^2)^2\big) + h_0 \int_{-\infty}^\infty d\tau\, \phi_N (\tau,\mathbf{0})\,,
\end{equation}
where $a=1,\ldots,N-1$. The bulk action has $O(N)$ symmetry. This is broken to $O(N-1)$ by the choice of $\phi_N$ in the last term, which defines the line defect along $x^\mu(\tau)=(\tau,\mathbf{0})$. In the following we omit the index $N$ from $\phi_N$.

Standard renormalisation in the bulk results in the definition of a renormalised coupling $\lambda$ with beta function
\begin{equation}\label{eq:betalam}
    \beta_\lambda=-\varepsilon\lambda+\frac{1}{16\pi^2}\frac{N+8}{3}\lambda^2+\text{O}(\lambda^3)\,.
\end{equation}
This is not affected by the presence of the defect deformation. The non-trivial zero $\lambda_*$ of \eqref{eq:betalam} defines the $O(N)$ symmetric CFT in $d=4-\varepsilon$. Then one can consider the line deformation in \eqref{eq:action} and a renormalisation group flow localised on the line is triggered due to the beta function in \eqref{eq:beta_h} for the renormalised coupling $h$. The non-trivial zero of $\beta_h$ defines a dCFT, i.e.\ a one-dimensional CFT localised on the line defect.

In this work we perform detailed analytic bootstrap studies of four-point functions in the $O(N)$ magnetic line dCFT, and it is useful to start by summarising the form that two-, three- and four-point functions take. In general, two-point functions of renormalised primary operators are given by 
\begin{equation}\label{eq:two_PF}
    \langle \mathcal{O}_I (\tau) \mathcal{O}_J (0) \rangle = \frac{\mathcal{T}_{IJ}}{\tau^{2\Delta_I}}\,.
\end{equation}
Here labels $I,J$ are generalised indices that include the operators and their quantum numbers under the global symmetries $O(N-1)$ and $O(d-1)$. Then $\mathcal{T}_{IJ}$ identifies the primaries and has the appropriate group tensor structures. The scaling dimension of $\mathcal{O}_I$ is denoted by $\Delta_I$.

The three-point functions take the form
\begin{equation}\label{eq:three_PF}
    \langle \mathcal{O}_I (\tau_1) \mathcal{O}_J (\tau_2) \mathcal{O}_K (\tau_3) \rangle = \frac{ \lambda_{IJK}}{\lvert \tau_{12} |^{\Delta_I+\Delta_J-\Delta_K} \lvert \tau_{13} |^{\Delta_I-\Delta_J+\Delta_K} \lvert \tau_{23} |^{-\Delta_I+\Delta_J+\Delta_K}}\,,
\end{equation}
where $\lambda_{IJK}$ is the three-point function coefficient and $\tau_{ij}=\tau_i-\tau_j$. We also have the OPE
\begin{equation}
    \mathcal{O}_I(\tau)\times\mathcal{O}_J(0)= \frac{1}{|\tau|^{\Delta_I+\Delta_J-\Delta_K}}\lambda_{IJ}{\lnsp}^K\mathcal{O}_K(0)+\cdots,
\end{equation}
and consistency with \eqref{eq:two_PF} and \eqref{eq:three_PF} demands
\begin{equation}
    \lambda_{IJK}=\lambda_{IJ}{\lnsp}^L\mathcal{T}_{LK}\,.
\end{equation}

Four-point functions of primary operators take the form 
\begin{equation}\label{eq:gen4PF}
    \langle \mathcal{O}_I(\tau_1) \mathcal{O}_J(\tau_2) \mathcal{O}_K(\tau_3) \mathcal{O}_L(\tau_4) \rangle = \frac{1}{\left| \tau_{12} \right|^{\Delta_I+\Delta_J}\left| \tau_{34} \right|^{\Delta_K+\Delta_L}}\left( \frac{\left|\tau_{24}\right|}{\left|\tau_{14}\right|} \right)^{\Delta_I-\Delta_J}\left( \frac{\left|\tau_{14}\right|}{\left|\tau_{13}\right|} \right)^{\Delta_K-\Delta_L}G_{IJKL}(\chi)\,,
\end{equation}
where 
\begin{equation}
    \chi=\frac{\tau_{12}\tau_{34}}{\tau_{13}\tau_{24}}
\end{equation}
is the conformally invariant cross-ratio. If we order the operators along the defect as $\tau_1<\tau_2<\tau_3<\tau_4$, then the cross-ratio satisfies $0<\chi<1$. The function $G_{IJKL}(\chi)$ can be expanded in s-channel conformal blocks as
\begin{equation}\label{eq:conformal_block_decomp} 
    G_{IJKL}(\chi)= \sum_{M,N} \mathcal{T}^{MN}\lambda_{IJM}\lambda_{KLN}\, g_{\Delta_M}^{\Delta_{IJ},\Delta_{KL}}(\chi)\,,
\end{equation}
where $\mathcal{T}^{IJ}\mathcal{T}_{JK}=\delta^I{\!}_K$ and
\begin{equation}\label{block}
    g_{\Delta_M}^{\Delta_{IJ},\Delta_{KL}}(\chi)=\chi^{\Delta_M} {_{2}F_{1}}(\Delta_M-\Delta_{IJ},\Delta_M+\Delta_{KL};2\Delta_M;\chi)\,,
\end{equation}
where $\Delta_{IJ}=\Delta_I-\Delta_J$ and similarly for $\Delta_{KL}$ and ${_{2}F_{1}}$ is a hypergeomentric function. For our purposes below we split the sum over $M$ in \eqref{eq:conformal_block_decomp} into subspaces of operators ${\cal M}_{n,s,J}$ of fixed quantum numbers $s$ under $O(d-1)$, $J$ under $O(N-1)$ and order $\varepsilon^0$ dimension $n\in\mathbb{N}$. Thus,
\begin{equation}\label{eq:conformal_block_decomp_3} 
G_{IJKL}(\chi)= \sum_{n,s,J} \mathcal{T}^{(n,s,J)}_{IJKL}
\sum_{M\in\cM_{n,s,J}} 
\cG_{M}(\chi)\,,
\end{equation}
where $\cT^{(n,s,J)}_{IJKL}$ is the projector on the invariant subspaces of the four-point function and
\begin{equation}\label{dimension block} 
    \cG_M(\chi)=\frac{1}{\mathcal{T}_M}\lambda_M^2\, g_{\Delta_M}^{\Delta_{ij},\Delta_{kl}}(\chi)\,.
\end{equation}
With these steps we have effectively reorganised the sum over primaries in \eqref{eq:conformal_block_decomp} into a sum over primaries at each symmetry subspace and classical dimension. Note that $\lambda_M$ here are three-point function coefficients and $1/\mathcal{T}_M$ is found from the normalisation in \eqref{eq:two_PF}.

We largely assume unit normalised operators below so that there is no difference between OPE and three-point function coefficients. The tilt operator $t_a$ and displacement operator $D_\mu$, which have integer dimensions and canonical normalisations inherited from the bulk $O(N)$ current and energy momentum tensor, respectively, are special cases. For simplicity we still rescale them to be unit normalised except in Section~\ref{sec:integrals}, where we revert to the canonical normalisations to write the integral identities.

\section{Crude analytic bootstrap and unmixing}
\label{sec:analytic}

The four-point function of $\phi$ (the singlet defect operator coming from the field $\phi_N$) at linear order in $\varepsilon$ was calculated in \cite{Gimenez-Grau:2022czc}. The associated function $G(\chi)$ that features in \eqref{eq:gen4PF} in this case is given by
\begin{equation}\label{phiphiphiphi}
    G(\chi)=1+\chi^{2\Delta_{\phi}}+\Big(\frac{\chi}{1-\chi}\Big)^{2\Delta_{\phi}}+\frac{6}{N+8} I_1(\chi)\lsp\varepsilon+\text{O}(\varepsilon^2)\,,
\end{equation}
with
\begin{equation}
    I_1(\chi)=\chi\log(1-\chi)+\frac{\chi^2}{1-\chi}\log\chi\,.
\end{equation}
Explicitly, expanding \eqref{phiphiphiphi} in powers of $\varepsilon$ and $\chi$ and 
using $\Delta_\phi=1+\varepsilon+\text{O}(\varepsilon^2)$, we find
\begin{align}
G(\chi)
&=1
+\chi^2\left(2-\frac{6}{N+8}\lsp\varepsilon+\frac{2(2N+19)}{N+8}\lsp\varepsilon\log\chi\right)
+\chi^3\left(2+\frac{2N+13}{N+8}\lsp\varepsilon+\frac{2(2N+19)}{N+8}\lsp\varepsilon\log\chi\right)
\nonumber\\&\quad
+\chi^4\left(3+\frac{5N+38}{N+8}\lsp\varepsilon+\frac{6(N+9)}{N+8}\lsp\varepsilon\log\chi\right)
+\text{O}(\chi^5, \varepsilon^2)\,.
\label{Gpppp-expanded}
\end{align}
Here we have only a singlet sector to sum over in \eqref{eq:conformal_block_decomp_3}. The 1 comes from the identity operator and the $\chi^2$ terms match with an operator of dimension close to two. To see that, we expand the hypergeometric in \eqref{block}, where for an operator $M$ of dimension close to $n$,
\begin{equation}
    \Delta_{M}=n+\varepsilon\gamma_M+\text{O}(\varepsilon^2)\,,
\end{equation}
$\cG_M$ in \eqref{dimension block} is
\begin{align}\label{eq:CB-ep-expansion}
    \mathcal{G}_{n}(\chi)=\lambda_n^2\chi^n(1+\varepsilon\gamma_n\log\chi)(1+\text{O}(\chi,\varepsilon))\,.
\end{align}
Comparing to the $\chi^2$ term in \eqref{Gpppp-expanded} and assuming that a single operator of dimension close to two is exchanged, we find
\begin{equation}\label{dim2mix}
    \lambda_{2}^2=2-\frac{6}{N+8}\lsp\varepsilon+\text{O}(\varepsilon^2)\,,\qquad
    \lambda_{2}^2\gamma_2=\frac{2(2N+19)}{N+8}+\text{O}(\varepsilon)\quad\Rightarrow\quad\gamma_2=\frac{2N+19}{N+8}+\text{O}(\varepsilon)\,.
\end{equation}

In fact, if we expand the conformal block \eqref{block} to higher orders in $\chi$, the $\chi^3$ term in \eqref{Gpppp-expanded} is exactly matched, meaning that no operators of dimension close to three are exchanged at this order in $\varepsilon$. Then the $\chi^4$ term provides information on the exchanged operators of dimension close to four. Continuing these expansions to order $\chi^{40}$, we find data consistent with the general formulas
\begin{align}\label{eq:phi-gamma}
    \gamma_{n}&=2+\frac{6}{n(n-1)(N+8)}+\text{O}(\varepsilon)\,,\\
    \label{eq:phi-lambda}
    \lambda_n^2&=\begin{cases}    
        \text{O}(\varepsilon^2)\,,&\text{$n$ odd,}\\[2pt]\displaystyle\frac{n!(n-1)!}{(2n-3)!}\big(1-\big(4-4H_n+2\gamma_n(H_{2n-2}-H_{n-1})\big)\varepsilon\big)+\text{O}(\varepsilon^2)\,,&\text{$n$ even,}
    \end{cases}
\end{align}
where $H_n$ is the $n$-th harmonic number. This matches the expressions found in \cite{Dey:2024ilw} using Polyakov bootstrap techniques.

The four-point functions of tilts and the mixed $\phi$-tilt four-point function were also calculated in \cite{Gimenez-Grau:2022czc}. The function $G_{abcd}(\chi)$ appearing in $\langle t_at_bt_ct_d\rangle$ is \cite{Gimenez-Grau:2022czc}
\begin{equation}\label{Gtttt}
    \begin{aligned}
        G_{abcd}(\chi)&=\delta_{ab}\delta_{cd}\lsp G^S(\chi)+\frac12\Big(\delta_{ac}\delta_{bd}+\delta_{ad}\delta_{bc}-\frac{2}{N-1}\delta_{ab}\delta_{cd}\Big)G^T(\chi)+(\delta_{ac}\delta_{bd}-\delta_{ad}\delta_{bc})G^A(\chi)\\
        &=\delta_{ab}\delta_{cd}+\delta_{ac}\delta_{bd}\lsp\chi^2+\delta_{ad}\delta_{bc}\Big(\frac{\chi}{1-\chi}\Big)^2+\frac{6}{N+8}(\delta_{ab}\delta_{cd}+\delta_{ac}\delta_{bd}+\delta_{ad}\delta_{bc}) I_1(\chi)\lsp\varepsilon+\text{O}(\varepsilon^2)\,.
    \end{aligned}
\end{equation}
Repeating the analysis described above for the singlet channel $G^S$ of the tilt four-point function, we again can fit the expansion using a single operator at every integer dimension, but now with
\begin{align}\label{eq:gamma-tilt-s}
    \gamma_{n}&=\frac{2(N+1)}{n(n-1)(N+8)}+\text{O}(\varepsilon)\,,\\
    \label{eq:lambda-tilt-s}
    \lambda_n^2&=\begin{cases}
        \text{O}(\varepsilon^2)\,,&\text{$n$ odd,}\\[2pt]
        \displaystyle\frac{n!(n-1)!}{(2n-3)!(N-1)}\big(1-2\gamma_n(H_{2n-2}-H_{n-1})\varepsilon\big)+\text{O}(\varepsilon^2)\,,&\text{$n$ even.}
    \end{cases}
\end{align}
These expressions are different from \eqref{eq:phi-gamma}, \eqref{eq:phi-lambda}.

The function $G_{ab}(\chi)$ in the expression for $\langle \phi\phi t_a t_b\rangle$ is given by
\begin{equation}\label{Gphiphitt}
    G_{ab}(\chi)=\delta_{ab}\Big[1+\frac{2}{N+8} I_1(\chi)\lsp\varepsilon +\text{O}(\varepsilon^2)\Big]\,.
\end{equation}
For this function, a procedure like the one described above is now unsuccessful if a single operator is assumed to be exchanged for each $n\geq2$. Indeed, assuming the exchange of a single operator $s$ of dimension close to two leads to 
\begin{equation}
\lambda_{\phi\phi s} \lambda_{t t s} = 0\,,\qquad
\gamma_s \lambda_{\phi\phi s} \lambda_{t t s} = \frac{2}{N+8}\,.
\end{equation}
Clearly these equations are inconsistent and this indicates that there are in fact multiple physical operators that are exchanged.
Assuming the exchange of multiple operators $s_i$ has the effect of replacing $\lambda_{\phi\phi s} \lambda_{t t s} \rightarrow \sum_i \lambda_{\phi\phi s_i} \lambda_{t t s_i}$ and likewise for $\gamma_s \lambda_{\phi\phi s} \lambda_{t t s}$, where the index $i$ runs over the exchanged operators.
This tells us that the results in \eqref{eq:phi-gamma}, \eqref{eq:phi-lambda}, \eqref{eq:gamma-tilt-s} and \eqref{eq:lambda-tilt-s} represent averages over the contributions of multiple operators.
In principle, we can assume the exchange of any arbitrarily large number of operators, but equations of this form can be made to be consistent by assuming the exchange of just two.

This is not surprising, as the number of physical scalars with dimension close to two is two. For $n>2$, there are more scalars; still, assuming two operators at each even $n$ is enough to simultaneously bootstrap the singlet channels of all three correlators. For $n=2$, this bootstrap completely resolves the spectrum of singlets. So we turn now from examining single correlation functions to systems of multiple ones to focus on full sectors of exchanged operators in \eqref{eq:conformal_block_decomp_3}, starting with $\cM_{2,\text{singlet},\text{singlet}}$.

As mentioned, to unmix this system it suffices to look at the matrix of correlation functions restricted to the singlet operator exchange 
\begin{equation}\label{pptt2x2}
    \begin{pmatrix}
        G&G_{ab}\\
        G_{ab}&G_{abcd}
    \end{pmatrix}_\text{singlet}=\cG_0+\cG_2+\cdots.
\end{equation}
Explicit computation as above gives
\begin{equation}\label{eq:G0G2mixed}
    \begin{aligned}
        \cG_0&=\begin{pmatrix}
            1&1\\
            1&1
        \end{pmatrix},\\
    \cG_2&=\chi^2\begin{pmatrix}
        2&0\\
        0&\frac{2}{N-1}
    \end{pmatrix}+\frac{2}{N+8}\lsp\chi^2\begin{pmatrix} 
        -3+(2N+19)\log\chi&-1+\log\chi\\
        -1+\log\chi&-\frac{N+1}{N-1}(1-\log\chi)
    \end{pmatrix}\varepsilon+\text{O}(\chi^3)\,.
    \end{aligned}
\end{equation}
Now, assuming two exchanged operators, $s_\pm$, there are four structure constants and two anomalous dimensions that we arrange in the $2\times2$ matrices
\begin{equation}\label{LG}
    \Lambda=\begin{pmatrix}
        \lambda_{\phi\phi s_+}&\lambda_{\phi\phi s_-}\\
        \lambda_{tt s_+}&\lambda_{tt s_-}
    \end{pmatrix},
    \qquad
    \Gamma=\begin{pmatrix}
        \gamma_{s_+}&0\\
        0&\gamma_{s_-}
    \end{pmatrix}.
\end{equation}
Then, to order $\varepsilon$ we may express
\begin{equation}\label{block 2x2}
    \cG_2 = \Lambda\Lambda^T\chi^2 + \varepsilon\lsp \Lambda\Gamma\Lambda^T\chi^2\log\chi+\text{O}(\chi^3,\varepsilon^2)\,,
\end{equation}
and read off
\begin{equation}\label{LL2-1}
    \Lambda\Lambda^T=\begin{pmatrix}
        2&0\\
        0&\frac{2}{N-1}
    \end{pmatrix}
    -\frac{2}{N+8}\begin{pmatrix}
        3&1\\
        1&\frac{N+1}{N-1}
    \end{pmatrix}\varepsilon\,,
    \qquad
    \Lambda\Gamma\Lambda^T
    =\frac{2}{N+8}\begin{pmatrix}
        2N+19&1\\
        1&\frac{N+1}{N-1}
    \end{pmatrix},
\end{equation}
on comparison with \eqref{eq:G0G2mixed}. The anomalous dimensions are the eigenvalues of the matrix
\begin{equation}\label{nondiag}
    \Lambda\Gamma\Lambda^T(\Lambda\Lambda^T)^{-1}=\Lambda\Gamma\Lambda^{-1}=
    \frac{1}{N+8}\begin{pmatrix}
        2N+19&N-1\\
        1&N+1
    \end{pmatrix}
    +\text{O}(\varepsilon)\,,
\end{equation}
giving
\begin{equation}
    \gamma_{s_\pm}=\frac{3N+20\pm\sqrt{N^2+40N+320}}{2(N+8)}\,,
\end{equation}
which are in accord with the results of~\cite{Gimenez-Grau:2022czc}.

The operators $s_\pm$ appear in other OPEs as well. For example, consider the operator $V_a\sim\phi\phi_a$ of classical scaling dimension close to two. We can do the calculation outlined above replacing $\langle\phi\phi t_at_b\rangle$ with $\langle\phi\phi V_a V_b\rangle$ and $\langle t_at_bt_ct_d\rangle$ with $\langle V_aV_bV_cV_d\rangle$. The four-point functions $\langle\phi\phi V_aV_b\rangle$ and $\langle V_aV_bV_cV_d\rangle$ can be found in the ancillary \emph{Mathematica} file. From those we extract
\begin{equation}\label{LL2-2}
    \Lambda\Lambda^T= \begin{pmatrix}
        2 & 2 \\
        2 &\frac{2N}{N-1}
    \end{pmatrix}+\text{O}(\varepsilon)\,,\qquad
    \Lambda\Gamma\Lambda^T=\frac{2}{N+8}\begin{pmatrix}
        2N+19&2(N+10)\\2(N+10)& \frac{2(N^2+10N-10)}{N-1}
    \end{pmatrix}.
\end{equation}
The product $\Lambda\Gamma\Lambda^T (\Lambda\Lambda^T)^{-1}$ is a different matrix from \eqref{nondiag}, but obviously has the same eigenvalues as above.

Finally, we can perform a similar computation using $\langle t_at_bt_ct_d\rangle$, $\langle t_at_bV_cV_d\rangle$ and $\langle V_aV_bV_cV_d\rangle$. This results in
\begin{equation}\label{LL2-3}
    \Lambda\Lambda^T= \begin{pmatrix}
        \frac{2}{N-1}&\frac{2}{N-1}\\
        \frac{2}{N-1}&\frac{2N}{N-1}
    \end{pmatrix}+\text{O}(\varepsilon)\,,\qquad
    \Lambda\Gamma\Lambda^T=\frac{2}{N+8}\begin{pmatrix}
        \frac{N+1}{N-1} &\frac{2N}{N-1}\\
        \frac{2N}{N-1} &\frac{2(N^2+10N-10)}{N-1} 
    \end{pmatrix},
\end{equation}
and these give the same two eigenvalues as well.

The six correlation functions of pairs of $\phi$, $t_a$ and $V_a$ can be arranged into a larger (symmetric) $3\times3$ system of correlation functions, which to leading order is determined by
\begin{align}\label{LL3}
    \Lambda\Lambda^T= \begin{pmatrix}
        2 & 0 & 2 \\
        0 & \frac{2}{N-1} & \frac{2}{N-1} \\
        2 & \frac{2}{N-1} & \frac{2N}{N-1}
    \end{pmatrix},\qquad\Lambda\Gamma\Lambda^T=\frac{2}{N+8} \begin{pmatrix}
        2N+19& 1& 2(N+10)\\
        1& \frac{N+1}{N-1} & \frac{2N}{N-1} \\
        2(N+10)& \frac{2N}{N-1} & \frac{2(N^2+10N-10)}{N-1}
    \end{pmatrix}.
\end{align}
The three previous $2\times2$ systems are the simultaneous submatrices we get by removing one row and the corresponding column from both matrices. Crucially, both $3\times3$ matrices are degenerate, but we find the same eigenvalues for $\Lambda\Gamma\Lambda^T(\Lambda\Lambda^T)^{-1}$ from each of the three subsystems.

To understand the properties of these matrices and their submatrices, consider $M_1=\Lambda\Lambda^T$ a $k\times k$ matrix of rank $l$. It is symmetric and there exists an $l\times k$ matrix $\cal R$ such that $\cR M_1\cR ^T=\mathds{1}_l$, the $l\times l$ identity matrix. One can also replace $\Lambda$ with a $k\times l$ matrix $L$ such that $M_1=LL^T$ and $\cR L=\mathds{1}_l$. Then we can rewrite $M_2=\Lambda\Gamma\Lambda^T=LDL^T$ with $D=\cR M_2\cR ^T$ and clearly the eigenvalues of $\Gamma$ are the $l$ eigenvalues of $D$ and some extra eigenvalue that plays no role in $M_2$ and can be assumed to vanish. While this procedure works, in practice we actually do not need to find this particular $\cR$. Instead we take any $l\times k$ matrix $R$ of rank $l$ such that the image of $R^T$ does not include the kernel of $\Lambda^T$. Under these conditions, $RM_1R^T=RL(RL)^T$ is an invertible $l\times l$ matrix and $(RM_2R^T)(RM_1R^T)^{-1}=(RL)D(RL)^{-1}$ is a full rank $l\times l$ matrix with the same eigenvalues as $RM_2R$, or the relevant eigenvalues of $\Gamma$. In the example above, $R$ is any of the $2\times 3$ matrices $R_i$ with $i=1,2,3$ defined such that their left action on a $3\times 3$ matrix removes the $i$-th row. In Section~\ref{sec:spectrum} we sometimes need to remove more than a single line.

Let us now discuss the evaluation of three-point functions or OPE coefficients. $\Lambda$ appears in both $\Lambda\Lambda^T$ and $\Lambda\Gamma\Lambda^T$ and assuming they are $k\times k$ and regular, their determination gives $k(k+1)$ quantities, which are the $k$ eigenvalues of $\Gamma$ and the $k^2$ entries in $\lambda$ (up to signs, since the equations are quadratic). However, our determination of these matrices is only perturbative and limited to $\text{O}(\varepsilon)$. Since $\Gamma$ already accounts for one power of $\varepsilon$, $\Lambda\Gamma\Lambda^T$ can only help to determine $\Lambda$ at $\text{O}(\varepsilon^0)$, which with information from $\Lambda\Lambda^T$ is completely determined. The $\text{O}(\varepsilon)$ part of $\Lambda$ appears only in $\Lambda\Lambda^T$ and constrained by $k(k+1)/2$ equations. One way to completely determine them is by evaluating the same set of four-point functions to order $\varepsilon^2$. Another is to consider further four-point functions, specifically ones where the roles of external and exchanged operators are interchanged.

For the example presented above, we follow the latter approach and calculate one extra four-point function, e.g.\ $\langle\phi s_-\phi s_-\rangle$, which allows then to determine all three-point function coefficients to order $\varepsilon$. We computed this four-point function using Feynman diagrams and applying our analytic bootstrap procedure we find an exchanged operator of dimension close to one, which is simply $\phi$ in the $\phi\times s_-$ OPE. This gives us explicitly $\lambda_{\phi s_-\phi}^2=\lambda_{\phi\phi s_-}^2$ and with this information as well as the information contained in $\langle\phi\phi\phi\phi\rangle$, $\langle tttt\rangle$ and $\langle\phi\phi tt\rangle$, we obtain
\bal
    |\lambda_{\phi\phi s_\pm}|&=\sqrt{1\pm\frac{N+18}{\sqrt{N^2+40N+320}}}\bigg(1+\frac{N+12\mp\sqrt{N^2+40N+320}}{4(N+8)}\varepsilon+\text{O}(\varepsilon^2)\bigg)\,,\\
    |\lambda_{tt s_\pm}|&=\frac{1}{\sqrt{N-1}}\sqrt{1\mp\frac{N+18}{\sqrt{N^2+40N+320}}}\bigg(1-\frac{3N+20\pm\sqrt{N^2+40N+320}}{4(N+8)}\varepsilon+\text{O}(\varepsilon^2)\bigg)\,.
\label{lambdaep}
\eal
The $\text{O}(\varepsilon)$ part for these three-point function coefficients disagree with \cite[Eqs.\ (3.53), (3.54)]{Gimenez-Grau:2022czc}. Note that for $N=1$, a case for which there is no tilt operator $t$, there is only one of the $s_\pm$ operators that survives and this is reflected in the fact that $|\lambda_{\phi\phi s_-}|=0$ for $N=1$.

\section{The spectrum}
\label{sec:spectrum}

In this section we collect the results for the spectrum of defect operators that we 
evaluated.

\subsection{Dimension \texorpdfstring{$\Delta\sim1$}{Delta~1}}\label{sec:dim1}
At dimension close to one we have the operators made of the basic fields $\phi_a$ and $\phi$, both acquiring anomalous dimensions compared to the free-field value of $1-\varepsilon/2$~\cite{Cuomo:2021kfm, Gimenez-Grau:2022czc}. The former are known as tilts and play an important role in the integral identities in Section~\ref{sec:integrals} below. The details including their representation under $O(N-1)$ and the transverse 
rotation group $O(d-1)$ are in Table~\ref{tab:dim1}.
\begin{table}[ht]
\centering
\begin{tabular}{ |l|c|c|l|l| } 
\hline
Name&$O(N-1)$& $O(d-1)$&Dimension&Fields \\\hline
$\phi$&$\boldsymbol{1}$&$\boldsymbol{1}$&$1+\varepsilon$&$\phi$\\\hline
$t_a$&$\boldsymbol{N-1}$&$\boldsymbol{1}$&1&$\phi_a$\\\hline
\end{tabular}
\caption{Operators of dimension near 1\label{tab:dim1}}
\end{table}

\subsection{Dimension \texorpdfstring{$\Delta\sim2$}{Delta~2}}
\label{sec:dim2}
The primary operators of dimension close to two are listed in Table~\ref{tab:dim2}. All of them were already studied in \cite{Gimenez-Grau:2022czc} and we rederived them using Feynman diagrams. We find the dimensions and structure constants of $s_\pm$ also from the expansion of the four-point functions of operators of dimension close to one as explained in Section~\ref{sec:analytic}. Note, though, that this process does not provide the realisation of $s_\pm$ in terms of the free fields, which is required to calculate further four-point functions and unmix higher dimensional operators.

The dimensions of $V_a$ and $T_{ab}$ can also be determined from four-point functions, as already shown in~\cite{Gimenez-Grau:2022czc}. The dimension of $D_\mu$ is fixed to two, as it is the protected displacement operator.

\begin{table}[ht]
\centering
\begin{tabular}{ |l|c|c|l|l| } 
\hline
Name&$O(N-1)$& $O(d-1)$&Dimension&Fields \\\hline
$s_+$, $s_-$&$\boldsymbol{1}$&$\boldsymbol{1}$ &
$2+\frac{3N+20\pm\sqrt{N^2+40N+320}}{2(N+8)}\varepsilon$&
$\phi\phi$, $\phi_a\phi_a$\\\hline
$V_a$&$\boldsymbol{N-1}$&$\boldsymbol{1}$&
$2+\frac{N+10}{N+8}\varepsilon$
&$\phi\phi_a$\\\hline
$T_{ab}$&$\boldsymbol{\frac{(N+1)(N-2)}{2}}$&$\boldsymbol{1}$&
$2+\frac{2}{N+8}\varepsilon$
&$\phi_a\phi_b-\tr$\\\hline
$D_\mu$&$\boldsymbol{1}$&$\boldsymbol{d-1}$&2&$\partial_\mu\phi$\\\hline
$U_{a\mu}$&$\boldsymbol{N-1}$&$\boldsymbol{d-1}$&$2-\frac{1}{3}\varepsilon$&$\partial_\mu\phi_a$\\\hline
\end{tabular}
\caption{Operators of dimension near 2\label{tab:dim2}}
\end{table}

\subsection{Dimension \texorpdfstring{$\Delta\sim3$}{Delta~3}}
\label{sec:dim3}

One of the main results of this work is the full spectrum of defect primaries of 
dimension close to 3, as listed in Table~\ref{tab:dim3}. They were mostly found 
via the algorithm explained in Section~\ref{sec:analytic}. Note that we do not include operators that are removed by the equation of motion.
\begin{table}[ht!]
\centering
\begin{tabular}{ |l|c|c|l|l| } 
\hline
Name&$O(N-1)$& $O(d-1)$&Dimension&Fields \\\hline
$s'_\pm$,&
$\boldsymbol{1}$&
$\boldsymbol{1}$&
$3+\frac{5N+46\pm\sqrt{N^2+52N+388}}{2(N+8)}\varepsilon$&
$\phi\phi\phi$, $\phi_a\phi_a\phi$, \\\hline
$V'_{\pm,a}$,&
$\boldsymbol{N-1}$&
$\boldsymbol{1}$&
$3+\frac{3N+30\pm\sqrt{N^2+36N+260}}{2(N+8)}\varepsilon$
&$\phi\phi\phi_a$, $\phi_b\phi_b\phi_a$,\\\hline
$P_{a}$&$\boldsymbol{N-1}$&
$\boldsymbol{1}$&
$3+\varepsilon$&$\phi\partial_\tau\phi_a$\\\hline
$A_{ab}$&$\boldsymbol{\frac{(N-1)(N-2)}{2}}$&$\boldsymbol{1}$&3&
$\phi_a\partial_\tau\phi_b-\phi_b\partial_\tau\phi_a$ 
\\\hline
$T'_{ab}$&$\boldsymbol{\frac{(N+1)(N-2)}{2}}$&$\boldsymbol{1}$&
$3+\frac{N+14}{N+8}\varepsilon$
&$\phi\phi_a\phi_b-\tr$\\\hline
$\cT_{abc}$&$\boldsymbol{\frac{(N+3)(N-1)(N-2)}{6}}$&$\boldsymbol{1}$&
$3+\frac{6}{N+8}\varepsilon$
&$\phi_a\phi_b\phi_c-\tr{}$\\\hline
$D'_{\pm\mu}$
&$\boldsymbol{1}$&$\boldsymbol{d-1}$&$3+\frac{5N+28\pm\sqrt{N^2+112N+1408}}{6(N+8)}\varepsilon$&
$\phi\partial_\mu\phi$, $\phi_a\partial_\mu\phi_a$\\\hline
$U'_{\pm,a\mu}$
&$\boldsymbol{N-1}$&$\boldsymbol{d-1}$&$3+\frac{N+11\pm\sqrt{N^2+16N+73}}{3(N+8)}\varepsilon$&$\phi\partial_\mu\phi_a$, 
$\phi_a\partial_\mu\phi$\\\hline
${\cal A}_{ab\mu}$&$\boldsymbol{\frac{(N-1)(N-2)}{2}}$&$\boldsymbol{d-1}$&$3-\frac{1}{3}\varepsilon$&
$\phi_a\partial_\mu\phi_b-\phi_b\partial_\mu\phi_a$\\\hline
$W_{ab\mu}$&$\boldsymbol{\frac{(N+1)(N-2)}{2}}$&$\boldsymbol{d-1}$&$3-\frac{N+2}{3(N+8)}\varepsilon$&
$\phi_a\partial_\mu\phi_b+\phi_b\partial_\mu\phi_a-\tr$\\\hline
${\cal D}_{\mu\nu}$&$\boldsymbol{1}$&$\boldsymbol{\frac{d(d-1)}{2}-1}$&
$3-\frac{1}{5}\varepsilon$&
$\partial_\mu\partial_\nu\phi-\tr$\\\hline
${\cal U}_{a\mu\nu}$&$\boldsymbol{N-1}$&$\boldsymbol{\frac{d(d-1)}{2}-1}$
&$3-\frac{2}{5}\varepsilon$&
$\partial_\mu\partial_\nu\phi_a-\tr$\\\hline
\end{tabular}
\caption{Operators of dimension near 3\label{tab:dim3}}
\end{table}

For the singlets at dimension $\Delta\sim3$, which in the language of \eqref{eq:conformal_block_decomp_3} is $\mathcal{M}_{3,S,S}$, the possible combinations of fields are $\phi\phi\phi$, $\phi_a\phi_a\phi$, $\partial_\mu \partial^\mu\phi$ and $\partial_\tau^2\phi$. The last one is a descendant of $\phi$ and the next to last is related to the first two operators by the equations of motion. To find the dimensions of the former two we note that they should appear in $\phi\times s_\pm$ and $t_a\times V_a$. The matrix of correlators which we consider (with repeated elements removed) is
\begin{equation}
    \begin{pmatrix}
        \langle \phi s_+ \phi s_+ \rangle & \langle \phi s_+ \phi s_- \rangle & \langle \phi s_+ t_a V_b \rangle \\
        & \langle \phi s_- \phi s_- \rangle & \langle \phi s_- t_a V_b \rangle \\
        && \langle t_a V_b t_c V_d \rangle_S
    \end{pmatrix}.
\end{equation}
The $3\times3$ system is degenerate in the sense that $\det\Lambda\Lambda^T=0$. Deleting any row and corresponding column leads to a non-degenerate $2\times2$ systems, which indicates that the full $3\times3$ system has rank two, as in the example in Section~\ref{sec:analytic}. This also supports the statements above regarding $\partial_\mu \partial^\mu\phi$ and $\partial_\tau^2\phi$.

For the case $\mathcal{M}_{3,S,V}$ of the $O(N-1)$ vectors, there are four fundamental field combinations, up to total $\tau$-derivatives, namely $\phi \phi \phi_a$, $\phi_b \phi_b \phi_a$, $\partial_\mu \partial^\mu \phi_a$ and $\phi \partial_\tau \phi_a$. The fourth actually decouples from the other three since it transforms differently under parity and unlike the others appears in OPE channels where one would normally look for even dimensional $O(N-1)$ vector operators. In particular, this operator makes a contribution to $\phi \times t_a$, $V_a \times s_\pm$ and $V_b \times T_{ba}$. We computed the correlators
\begin{equation}
    \begin{pmatrix}
        \langle \phi t_a \phi t_b \rangle & \langle \phi t_a V_b s_+ \rangle & \langle \phi t_a V_b s_- \rangle & \langle \phi t_a t_c T_{cb} \rangle \\
        & \langle V_a s_+ V_b s_+ \rangle & \langle V_a s_+ V_b s_- \rangle & \langle V_a s_+ t_c T_{cb} \rangle \\
        && \langle V_a s_- V_b s_- \rangle & \langle V_a s_- t_c T_{cb} \rangle \\
        &&& \langle t_c T_{ca} t_d T_{db} \rangle
    \end{pmatrix}.
\end{equation}
Naturally, anything larger than a $1\times1$ system is degenerate, showing that there really is only one exchanged operator of dimension close to three. Note that one needs to make sure here that the exchanged operator is properly taken to be $i\phi \partial_\tau \phi_a$ and not $\phi \partial_\tau \phi_a$. Otherwise, when looking at the matrix $\Lambda\Lambda^T$ for this system, the diagonal elements, which correspond to $\lambda_{\phi t \mathcal{O}}^2$, $\lambda_{Vs_\pm \mathcal{O}}^2$ and $\lambda_{VT\mathcal{O}}^2$ would be negative. This has been taken into account in the supplementary \emph{Mathematica} file.

The OPE channels that we use to probe the rest of this sector, which displays genuine operator mixing, are $\phi\times V_a$, $t_a\times s_\pm$ and $t_b \times T_{ba}$. The correlators that we need are
\begin{equation}
    \begin{pmatrix}
        \langle \phi V_a \phi V_b \rangle & \langle \phi V_a t_b s_+ \rangle & \langle \phi V_a t_b s_- \rangle & \langle \phi V_a t_c T_{cb} \rangle \\
        & \langle t_a s_+ t_b s_+ \rangle & \langle t_a s_+ t_b s_- \rangle & \langle t_a s_+ t_c T_{cb} \rangle \\
        && \langle t_a s_- t_b s_- \rangle & \langle t_a s_- t_c T_{cb} \rangle \\
        &&& \langle t_c T_{ca} t_d T_{db} \rangle
    \end{pmatrix}.
\end{equation}
This $4\times4$ system is degenerate, along with all possible $3\times3$ systems.
Any two rows and corresponding columns must be deleted to obtain a non-degenerate $2\times2$ system. 
This indicates that, once again, there are just two operators $V'_{\pm,a}$ that are involved in the mixing problem. As in the case of the singlets, the equations of motion eliminate the third potential operator by forcing it to be related to the other two and a descendant.

The antisymmetric operator $A_{ab}$ is the lone operator in $\mathcal{M}_{3,S,A}$ and is found in $t_a\times t_b$. The correlator to examine is $\langle t_a t_b t_c t_d \rangle_A$, as is done in \cite{Gimenez-Grau:2022czc}.

The two-index traceless-symmetric sector $\mathcal{M}_{3,S,T^{(2)}}$ is detected in $\phi\times T_{ab}$ and the traceless-symmetric channel of $t_a\times V_b$.
The relevant correlators are
\begin{equation}
    \begin{pmatrix}
        \langle t_a V_b t_c V_d \rangle_T & \langle t_a V_b \phi T_{cd} \rangle_T \\
        & \langle \phi T_{ab} \phi T_{cd} \rangle_T
    \end{pmatrix}.
\end{equation}
The $2\times2$ system is degenerate, while both $1\times1$ systems are non-degenerate. 
This is consistent with the fact that there is only one possible fundamental field realisation in this representation at dimension three, namely $\phi \phi_a \phi_b -\tr$. Since there is just a single operator $T'_{ab}$, we have enough constraints to determine the OPE coefficients to $\text{O}(\varepsilon)$ as
\bal
\label{lambdatVT'}
    \lambda_{tVT'} &= \sqrt{2}\left(1-\frac{2}{N+8}\varepsilon\right)\,,\qquad
    \lambda_{\phi TT'} &= 1-\frac{2}{N+8}\varepsilon\,.
\eal

Let us turn to sectors that have non-zero transverse spin under $O(d-1)$. There are two fundamental field combination that are $O(N-1)$ singlets and $O(d-1)$ vectors, namely $\phi \partial_\mu \phi$ and $\phi_a\partial_\mu\phi_a$. The corresponding operators belong to $\mathcal{M}_{3,V,S}$ and appear in the $O(N-1)$ singlet channels of $\phi\times D_\mu$ and $t_a \times U_{b\mu}$. The relevant correlators are
\begin{equation}
    \begin{pmatrix}
        \langle \phi D_\mu \phi D_\mu \rangle & \langle \phi D_\mu t_a U_{b\mu} \rangle \\
        & \langle t_a U_{b\mu} t_c U_{d\nu} \rangle_S
    \end{pmatrix}.
\end{equation}
This $2\times2$ system is non-degenerate, but as its dimension matches the 
explicit field combinations, this
$2\times2$ system is enough to fully unmix the sector 
to the two operators $D'_{\pm\mu}$.

For operators that transform as vectors under both $O(N-1)$ and $O(d-1)$, there are again two fundamental field realisations, $\phi\partial_\mu\phi_a$ and $\phi_a\partial_\mu\phi$. 
There is a linear combination of these that is a total derivative, but this is not a descendant since it is a transverse total derivative. 
The corresponding operators contribute to $\phi\times U_{a\mu}$ and $t_a \times D_\mu$, which means that the correlators to be considered are
\begin{equation}
    \begin{pmatrix}
        \langle \phi U_{a\mu} \phi U_{b\nu} \rangle & \langle \phi U_{a\mu} t_b D_\nu \rangle \\
        & \langle t_a D_\mu t_b D_\nu \rangle
    \end{pmatrix}.
\end{equation}
This is another $2\times2$ system which is non-degenerate but we can still 
fully unmix the system with 
just two operators $U'_{\pm,a\mu}$.

The remaining sectors are the antisymmetric and traceless-symmetric representations of $O(N-1)$ which also transform as vectors under $O(d-1)$. For both sectors, there is a single fundamental field combination, which stems from $\phi_a\partial_\mu\phi_b$ by either antisymmetrising the indices $a$ and $b$, or symmetrising and subtracting the trace. The corresponding operators appear in the antisymmetric and traceless-symmetric channels respectively of $t_a\times U_{b\mu}$. Therefore, these channels are the relevant components of the correlator $\langle t_a U_{b\mu} t_c U_{d\nu} \rangle$. We denote the antisymmetric operator by $A'_{ab\mu}$ and the traceless-symmetric operator by $W_{ab\mu}$. Since both of these sectors have a single operator, there is no mixing and we can easily read off their OPE coefficients to $\text{O}(\varepsilon)$ as
\bal
    \lambda_{tUW} & = 1-\frac{1}{2(N+8)}\varepsilon\,,\qquad
    \lambda_{tUA'} & = \frac{1}{\sqrt{2}}\,.
\eal

The remaining two sectors of operators with dimensions close to three both transform in the traceless-symmetric representation of $O(d-1)$, but transform as either a singlet $\mathcal{D}_{\mu\nu}$ or a vector $\mathcal{U}_{a\mu\nu}$ under $O(N-1)$. 
In terms of fundamental fields, these are given by $\partial_\mu \partial_\nu \phi -\tr$ and $\partial_\mu \partial_\nu \phi_a -\tr$ respectively. 
There is no OPE involving operators with dimensions close to one or two that contain contributions from these operators. They are therefore inaccessible to us through the use of our bootstrap methods.
Instead, we calculated the dimensions via Feynman diagrams and found them to be $\Delta_\mathcal{D}=3-\frac{1}{5}\varepsilon$ and $\Delta_\mathcal{U}=3-\frac{2}{5}\varepsilon$.

\subsection{Dimension \texorpdfstring{$\Delta\sim4$}{Delta~4}}
\label{sec:dim4}

\begin{table}[h!]
\centering
\begin{tabular}{ |l|c|c|l|l| } 
\hline
Name&$O(N-1)$& $O(d-1)$&Dimension&Fields \\\hline
\multirow{2}{*}{$s''_M$}&
\multirow{2}{*}{$\boldsymbol{1}$}&
\multirow{2}{*}{$\boldsymbol{1}$}&
\multirow{2}{*}{\eqref{dim4 singlet characteristic poly}}&
$\phi^4$, $(\phi_a\phi_a)^2$, $\phi^2(\phi_a)^2$, 
$\partial_\mu\phi\partial^\mu\phi$, 
\\&&&&
$\partial_\mu\phi_a\partial^\mu\phi_a$, 
$\phi\partial_\mu\partial^\mu\phi$, 
$\phi_a\partial_\mu\partial^\mu\phi_a$\\\hline
$p$&$\boldsymbol{1}$&
$\boldsymbol{1}$&
$4+\frac{2(N+5)}{N+8}\varepsilon$&$\phi_a\phi_a\partial_\tau\phi$\\\hline
$V''_{M,a}$&
$\boldsymbol{N-1}$&$\boldsymbol{1}$&$\eqref{dim4 vector dims}$&$\phi\phi\phi\phi_a$, $\phi_b\phi_b\phi\phi_a$, $\phi_a\partial_\mu \partial^\mu \phi$, $\phi\partial_\mu \partial^\mu \phi_a$\\\hline
$P'_{\pm,a}$
&$\boldsymbol{N-1}$&$\boldsymbol{1}$&\eqref{decoupled dim4 vector dims}&$\phi\phi\partial_\tau\phi_a$, $\phi_b\phi_b\partial_\tau\phi_a$\\\hline
\multirow{2}{*}{$\sigma^4_{M,ab}$}&
\multirow{2}{*}{$\boldsymbol{\frac{(N+1)(N-2)}{2}}$}&
\multirow{2}{*}{$\boldsymbol{1}$}&\multirow{2}{*}{$\eqref{dim4 2 index trsym dims}$}&
$(\phi\phi,\phi_d\phi_d)(\phi_a\phi_b-\tr)$\\
&&&&$\partial_\mu\phi_a\partial^\mu\phi_b-\tr$, $\phi_a\partial_\mu\partial^\mu\phi_b+(a\leftrightarrow b)-\tr$\\\hline
$\varpi_{ab}$&$\boldsymbol{\frac{(N+1)(N-2)}{2}}$&
$\boldsymbol{1}$&
$4+\frac{N+11}{N+8}\varepsilon$&$\phi_a\phi_b\partial_\tau\phi-\tr$\\\hline
$\cV^4_{abc}$&
$\boldsymbol{\frac{(N+3)(N-1)!}{6(N-3)!}}$&
$\boldsymbol{1}$&
$4+\frac{N+20}{N+8}\varepsilon$
&$\phi\phi_a\phi_b\phi_c-\tr$\\\hline
$\cT^4_{abcd}$&
$\boldsymbol{\frac{(N+5)N!}{24(N-3)!}}$&
$\boldsymbol{1}$&
$4+\frac{12}{N+8}\varepsilon$
&$\phi_a\phi_b\phi_c\phi_d-\tr{}$\\\hline
\end{tabular}
\caption{Some operators of dimension near 4
\label{tab:dim4}}
\end{table}

For the operators of dimension close to four, we unmix some specific sectors, including the singlet operators, of which there turns out to be eight. 
Possible fundamental field combinations are listed in the first two rows of Table~\ref{tab:dim4}. Among these field combinations, 
$\phi_a\phi_a \partial_\tau \phi$ is the only one with odd parity.
The corresponding operator $p$ appears in $\phi\times s_\pm$ and $t_a \times V_a$, which means that the relevant correlators are
\begin{equation}
    \begin{pmatrix}
        \langle \phi s_+ \phi s_+ \rangle & \langle \phi s_+ \phi s_- \rangle & \langle \phi s_+ t_a V_b \rangle \\
        & \langle \phi s_- \phi s_- \rangle & \langle \phi s_- t_a V_b \rangle \\
        && \langle t_a V_b t_c V_d \rangle_S
    \end{pmatrix}.
\end{equation}
The full $3\times3$ system is degenerate, along with all $2\times2$ systems obtained by deleting any row and corresponding column. 
Only the $1\times1$ systems are non-degenerate, confirming that just one operator of this dimension is exchanged.
We therefore can determine the OPE coefficients up to $\text{O}(\varepsilon)$ to be
\bal
    \lambda_{\phi s_\pm p} &= \frac{2}{\sqrt{6}}\sqrt{1\mp\frac{N+18}{\sqrt{N^2+40N+320}}} \left(1 +\frac{8N+82\pm3\sqrt{N^2+40N+320}}{12(N+8)}\varepsilon \right),\\
    \lambda_{t V p} &= \frac{1}{\sqrt{6(N-1)}}\left(2-\frac{N-34}{6(N+8)}\varepsilon \right).
\eal

The remaining even parity singlets appear in the singlet channels of the OPEs%
\footnote{Note that contracting $O(d-1)$ indices gives $\delta^\mu{\!}_\mu=d-1=3-\varepsilon$.}
\begin{equation}
\begin{aligned}\label{dim4 singlet OPEs}
&U_{a\mu}\times U_{b\nu}\,,\qquad
&&D_\mu\times D_\nu\,,\\
&\phi\times\phi\,,\qquad
&&t_a\times t_b\,,\\
&V_a\times V_b\,,\qquad
&&s_+\times s_+\,,\qquad
&&s_+\times s_-\,,\qquad
&&s_-\times s_-\,.
\end{aligned}
\end{equation}
It turns out that not all operators in this sector contribute to the first four OPEs in \eqref{dim4 singlet OPEs}. We therefore write the matrix of correlators in the form
\begin{equation}
    \begin{pmatrix}
        S_1 & S_3 \\ S_3^T & S_2
    \end{pmatrix},
\end{equation}
where
\begin{align}
    S_1 &= \begin{pmatrix}
        \langle U_{a\mu} U_{b\nu} U_{c\rho} U_{d\sigma} \rangle_S & \langle U_{a\mu} U_{b\nu} D_\rho D_\sigma \rangle_S & \langle U_{a\mu} U_{b\nu} \phi \phi \rangle & \langle U_{a\mu} U_{b\nu} t_c t_d \rangle_S \\
        & \langle D_\mu D_\nu D_\rho D_\sigma \rangle_S & \langle D_\mu D_\nu \phi \phi \rangle & \langle D_\mu D_\nu t_a t_b \rangle \\
        && \langle \phi \phi \phi \phi \rangle & \langle \phi \phi t_a t_b \rangle \\
        &&& \langle t_a t_b t_c t_d \rangle_S
    \end{pmatrix},\\
    S_2 &= \begin{pmatrix}
        \langle V_a V_b V_c V_d \rangle_S & \langle V_a V_b s_+ s_+ \rangle & \langle V_a V_b s_+ s_- \rangle & \langle V_a V_b s_- s_- \rangle \\
        & \langle s_+ s_+ s_+ s_+ \rangle & \langle s_+ s_+ s_+ s_- \rangle & \langle s_+ s_+ s_- s_- \rangle \\
        && \langle s_+ s_- s_+ s_- \rangle & \langle s_+ s_- s_- s_- \rangle \\
        &&& \langle s_- s_- s_- s_- \rangle
    \end{pmatrix},\\
    S_3 &= \begin{pmatrix}
        \langle U_{a\mu} U_{b\nu} V_c V_d \rangle_S & \langle U_{a\mu} U_{b\nu} s_+ s_+ \rangle & \langle U_{a\mu} U_{b\nu} s_+ s_- \rangle & \langle U_{a\mu} U_{b\nu} s_- s_- \rangle \\
        \langle D_{\mu} D_{\nu} V_c V_d \rangle_S & \langle D_{\mu} D_{\nu} s_+ s_+ \rangle & \langle D_{\mu} D_{\nu} s_+ s_- \rangle & \langle D_{\mu} D_{\nu} s_- s_- \rangle \\
        \langle \phi \phi V_a V_b \rangle & \langle \phi \phi s_+ s_+ \rangle & \langle \phi \phi s_+ s_- \rangle & \langle \phi \phi s_- s_- \rangle \\
        \langle t_a t_b V_c V_d \rangle_S & \langle t_a t_b s_+ s_+ \rangle & \langle t_a t_b s_+ s_- \rangle & \langle t_a t_b s_- s_- \rangle
    \end{pmatrix}.
\end{align}
The full $8\times8$ system is degenerate and removing any of the six OPEs from the second and third lines of \eqref{dim4 singlet OPEs} produces a regular $7\times 7$ system.
Removing either of the OPEs on the top line yields a degenerate $7\times 7$ system, while removing both produces a degenerate $6\times6$ system. 
Also, the OPEs in the top two lines together form a $4\times4$ subsystem in the sense that the four eigenvalues appear in the full non-degenerate $7\times7$ system. These four operators are a mixture of the fundamental fields with two derivatives
\begin{equation}
\partial_\mu\phi\partial^\mu\phi\,, \qquad
\partial_\mu\phi_a\partial^\mu\phi_a\,,\qquad
\partial_\tau\phi\partial_\tau\phi\,,\qquad
\partial_\tau\phi_a\partial_\tau\phi_a\,,
\end{equation}
(with descendants removed). The remaining three are mixtures that also include
\begin{equation}
\phi^4\,,\qquad
(\phi_a\phi_a)^2\,,\qquad
\phi^2(\phi_a)^2\,.
\end{equation}
The full characteristic polynomial of the $7\times7$ system factors into
\begin{equation}
\begin{aligned}
    \label{dim4 singlet characteristic poly} \left( z^4 -\frac{7N+44}{3(N+8)}z^3 +\frac{5N^2+6N-380}{9(N+8)^2}z^2+\frac{2(N^2+10N+100)}{9(N+8)^2}z-\frac{10(N+2)}{27(N+8)^2} \right) 
    \\\left( -z^3 +\frac{9N+92}{N+8}z^2-\frac{2(13N^2+254N+1248)}{(N+8)^2}z +\frac{24(N^2+20N+102)}{(N+8)^2} \right).
\end{aligned}
\end{equation}
The first factor is the characteristic polynomial of the $4\times4$ subsystem. The zeroes of this polynomial are the $\text{O}(\varepsilon)$ corrections to the dimensions of the operators $s''_M$.

The vector operators at dimension 4 tell a similar story to the singlet sectors. There are six fundamental field combinations, two of which have odd parity, i.e.\ $\phi \phi \partial_\tau \phi_a$ and $\phi_b\phi_b\partial_\tau\phi_a$. The operators corresponding to these two appear in $\phi\times V_a$, $t_a\times s_\pm$ and $t_b\times T_{ba}$. The related correlators are
\begin{equation}
    \begin{pmatrix}
        \langle \phi V_a \phi V_b \rangle & \langle \phi V_a t_b s_+ \rangle & \langle \phi V_a t_b s_- \rangle & \langle \phi V_a t_c T_{cb} \rangle \\
        & \langle t_a s_+ t_b s_+ \rangle & \langle t_a s_+ t_b s_- \rangle & \langle t_a s_+ t_c T_{cb} \rangle \\
        && \langle t_a s_- t_b s_- \rangle & \langle t_a s_- t_c T_{cb} \rangle \\
        &&& \langle t_c T_{ca} t_d T_{db} \rangle
    \end{pmatrix}.
\end{equation}
This $4\times4$ system is degenerate, along with all $3\times3$ subsystems.
The $2\times2$ systems are all non-degenerate, meaning that there are two operators $P'_{\pm,a}$ involved in the mixing, which agrees with the counting of free field combinations.
The dimensions of these operators are
\begin{equation}
    \label{decoupled dim4 vector dims} 4+\frac{3N+21\pm\sqrt{N^2+42N+353}}{2(N+8)}\varepsilon\,.
\end{equation}
The parity even vectors appear in the OPEs
\begin{equation}
\begin{aligned}\label{dim4 vec system}
    &D_\mu\times U_{a\nu}\,,\qquad &&\phi\times t_a\,,\\
    &V_a\times s_+\,,\qquad &&V_a\times s_-\,,\qquad &&V_b\times T_{ba}\,.
\end{aligned}
\end{equation}
The relevant correlators are
\begin{equation}
    \begin{pmatrix}
        \langle D_\mu U_{a\nu} D_\rho U_{b\sigma} \rangle_S & \langle D_\mu U_{a\nu} \phi t_b \rangle & \langle D_\mu U_{a\nu} V_b s_+ \rangle & \langle D_\mu U_{a\nu} V_b s_- \rangle & \langle D_\mu U_{a\nu} V_c T_{cb} \rangle \\
        & \langle \phi t_a \phi t_b \rangle & \langle \phi t_a V_b s_+ \rangle & \langle \phi t_a V_b s_- \rangle & \langle \phi t_a V_c T_{cb} \rangle \\
        && \langle V_a s_+ V_b s_+ \rangle & \langle V_a s_+ V_b s_- \rangle & \langle V_a s_+ V_c T_{cb} \rangle \\
        &&& \langle V_a s_- V_b s_- \rangle & \langle V_a s_- V_c T_{cb} \rangle \\
        &&&& \langle V_{c} T_{ca} V_d T_{db} \rangle
    \end{pmatrix}.
\end{equation}
While removing $D_\mu\times U_{a\nu}$ leads to a degenerate $4\times 4$ system, removing any other OPE leads to a non-degenerate $4\times4$ system. Moreover, the two OPEs in the top line of \eqref{dim4 vec system} form a $2\times2$ subsystem. The full characteristic polynomial factors as
\begin{equation}
    \left( z^2 -\frac{2(N+11)}{3(N+8)}z -\frac{3N+10}{9(N+8)} \right) \left( z^2 -\frac{5(N+12)}{(N+8)}z +\frac{6(N^2+23N+136)}{(N+8)^2} \right).
\end{equation}
The degree 2 polynomial in the first factor is the characteristic polynomial of the $2\times2$ subsystem. The zeroes of this polynomial are the $\text{O}(\varepsilon)$ corrections to the dimensions of the operators $V''_{M,a}$. Altogether, the four dimensions are
\begin{equation}
    \label{dim4 vector dims} 4+\frac{N+11\pm\sqrt{4N^2+56N+201}}{3(N+8)}\varepsilon\,,\qquad 4 +\frac{5N+60\pm\sqrt{N^2+48N+336}}{2(N+8)}\varepsilon\,.
\end{equation}

The final two sectors that we unmix using our bootstrap methods are the parity odd and parity even two-index $O(N-1)$ traceless-symmetric operators. 
There are five such fundamental field expressions, one of which has odd parity, $\phi_a\phi_b \partial_\tau \phi -\tr$. 
The corresponding operator appears in $\phi\times T_{ab}$ and $t_a\times V_b$, which means that the relevant correlators are
\begin{equation}
    \begin{pmatrix}
        \langle \phi T_{ab} \phi T_{cd} \rangle_T & \langle \phi T_{ab} t_c V_d \rangle_T \\
        & \langle t_a V_b t_c V_d \rangle_T
    \end{pmatrix}.
\end{equation}
This system is degenerate, which indicates that there is a single operator $\varpi_{ab}$ involved.
In addition to its dimension, which is written in Table \ref{tab:dim4}, it is also possible to determine its OPE coefficients to $\text{O}(\varepsilon)$. We find them to be
\bal
    \lambda_{\phi T \varpi} &= \frac{1}{\sqrt{3}} \left(2 + \frac{4N+29}{6(N+8)}\varepsilon\right),\qquad
    \lambda_{tV \varpi} &= \frac{1}{\sqrt{6}} \left( 2+\frac{4N+29}{6(N+8)}\varepsilon \right).
\eal
The parity even operators appear in the traceless-symmetric channels of the OPEs,
\begin{equation}
    U_{a\mu}\times U_{b\nu}\,, \qquad t_a \times t_b\,, \qquad V_a \times V_b\,, \qquad s_+ \times T_{ab}\,, \qquad s_- \times T_{ab}\,.
\end{equation}
The correlators that we use are
\begin{equation}
\label{sigma4}
    \begin{pmatrix}
        \langle U_{a\mu} U_{b\nu} U_{c\rho} U_{d\sigma} \rangle_{T;S} & \langle U_{a\mu} U_{b\nu} t_c t_d \rangle_T & \langle U_{a\mu} U_{b\nu} V_c V_d \rangle_T & \langle U_{a\mu} U_{b\nu} s_+ T_{cd} \rangle_T & \langle U_{a\mu} U_{b\nu} s_- T_{cd} \rangle_T \\
        & \langle t_a t_b t_c t_d \rangle_T & \langle t_a t_b V_c V_d \rangle_T & \langle t_a t_b s_+ T_{cd} \rangle_T & \langle t_a t_b s_- T_{cd} \rangle_T \\
        && \langle V_a V_b V_c V_d \rangle_T & \langle V_a V_b s_+ T_{cd} \rangle_T & \langle V_a V_b s_- T_{cd} \rangle_T \\
        &&& \langle s_+ T_{ab} s_+ T_{cd} \rangle_T & \langle s_+ T_{ab} s_- T_{cd} \rangle_T \\
        &&&& \langle s_- T_{ab} s_- T_{cd} \rangle_T
    \end{pmatrix},
\end{equation}
where $\langle U_{a\mu} U_{b\nu} U_{c\rho} U_{d\sigma} \rangle_{T;S}$ indicates the $O(N-1)$ traceless-symmetric $O(d-1)$ singlet channel of $\langle U_{a\mu} U_{b\nu} U_{c\rho} U_{d\sigma} \rangle$.
This $5\times5$ system is degenerate. 
If $U_{a\mu} \times U_{b\nu}$ is removed, then the resulting $4\times4$ system is also degenerate.
All other $4\times4$ systems are non-degenerate, indicating four operators $\sigma^4_{M,ab}$ involved in the mixing problem. Their dimensions are
\begin{equation}
    \label{dim4 2 index trsym dims} 
    4-\frac{N+5\pm\sqrt{N^2+12N+41}}{3(N+8)}\varepsilon\,,
    \qquad 
    4+\frac{3N+44\pm\sqrt{N^2+32N+208}}{2(N+8)}\varepsilon\,.
\end{equation}

The maximally symmetric traceless tensor and the one with three indices are both 
unique and don't undergo mixing. They fit the pattern in the next section 
of large charge operators.

\subsection{Large charge}
\label{sec:large-charge}

A sector that is relatively easy to study is that of the operators of largest $O(N-1)$ charge at a given dimension close to $J$. There is one operator with charge $J$ and one with $J-1$ and then more with $J-2$, listed in Table~\ref{tab:dimJ}. This is the pattern we see also for $\Delta\sim2,3,4$ in the subsections above, and as in those cases, there are more operators of charge $J-2$, which in terms of fundamental fields are obtained by acting on $\phi_{a_1}\cdots \phi_{a_{J-2}}$ with either $\partial_\tau^2$ or $\partial_\mu \partial^\mu$ before subtracting the trace terms.

\begin{table}[h!]
\centering
\begin{tabular}{ |l|c|c|l|l| } 
\hline
Name&$O(N-1)$& $O(d-1)$&Dimension&Fields \\\hline
$\cT^{J}_{a_1\cdots a_J}$&$\boldsymbol{\frac{(N+2J-3)(N+J-4)!}{J!(N-3)!}}$&$\boldsymbol{1}$&
$J+\frac{J(J-1)}{N+8}\varepsilon$
&$\phi_{a_1}\cdots \phi_{a_J}-\tr{}$\\\hline
$\cV^J_{a_1\cdots a_{J-1}}$&$\boldsymbol{\frac{(N+2J-5)(N+J-5)!}{(J-1)!(N-3)!}}$&$\boldsymbol{1}$&
$J+\varepsilon+\frac{J(J-1)}{N+8}\varepsilon$
&$\phi (\phi_{a_1}\cdots \phi_{a_{J-1}}-\tr{})$\\\hline
$\sigma^{J}_{M,a_1\cdots a_{J-2}}$&$\boldsymbol{\frac{(N+2J-7)(N+J-6)!}{(J-2)!(N-3)!}}$&$\boldsymbol{1}$&Not computed
&See $\sigma^4$ in Table~\ref{tab:dim4}\\
\hline
\end{tabular}
\caption{Large charge operators of dimension near $J$. We did not evaluate the dimensions of $\sigma^J_M$.\label{tab:dimJ}}
\end{table}

We label the operators with a superscript $J$ which is their approximate dimension, not their $O(N-1)$ quantum number. The charge can be seen in the number of indices on the operators in Table~\ref{tab:dimJ}, though we sometimes avoid the multiple indices by using a null polarisation vector $\theta^a$ and contract to 
\begin{equation}
\label{theta}
\cT^J(\tau,\theta)=\theta^{a_1} \cdots \theta^{a_J}\cT^J_{a_1\cdots a_J}\,.
\end{equation}
Note that for small $J$ the matching with the previous operators is 
$\cT^1_a\propto t_a$, $\cV^1=\phi$, $\cT^2_{ab}=T_{ab}$, $\cV^2_a=V_a$, 
$\sigma^2_\pm=s_\pm$ and so on.

Given that there is no mixing for $\cT^J$ and $\cV^J$, their dimensions are very 
easy to compute either by direct Feynman diagrams or from $t\times \cT^{J-1}$ and $t\times\cV^{J-1}$. The spectrum of $\sigma^J$ is complicated, as already there are four $\sigma^4_M$ operators. We do not pursue this further, though we do find some constraints on their CFT data in the next section.

\section{Integral identities}
\label{sec:integrals}

A further reason to look at the large charge sector is to use and exemplify another tool relating four-point functions to the spectrum, which is the integral identities of~\cite{Drukker:2022pxk} (see also 
\cite{Sakkas:2024dvm,Cavaglia:2022qpg, Herzog:2023dop}). As outlined in Section~\ref{sec:intro}, the identity \eqref{tilt-id} relates an integral of the connected part of the four-point function of tilt operators to the curvature of the space of defects, or the defect conformal manifold. The generalisation to the operators $\cT^J$ in the $J$-th symmetric traceless representation (c.f.\ Table~\ref{tab:dimJ}) is
\bal
\label{integral-id}
2 \int d\chi\log|\chi|& \langle t_a (1) t_b(\chi) \cT^J(0,\theta) \cT^J (\infty,\theta') \rangle_c
=(\theta_{a}\partial_{\theta^{b}} -\theta_{b}\partial_{\theta^{a}}) 
\langle \cT^J(0,\theta) \cT^J(\infty,\theta') \rangle\,.
\eal
The right-hand side is proportional 
to $J$, the charge of the operators, and for uncharged operators the 
right-hand side vanishes. Similar equations hold for other operators as well.

Equation \eqref{integral-id} as stated requires the canonical normalisation of the tilts, derived from their relation to the broken $O(N)$ currents, via $\partial^\tau J_\tau^{aN}+\partial^\mu J_\mu^{aN}=\delta(x_\perp)t^a$. In previous sections we rescaled it away, but here we reintroduce it, such that
\begin{equation}
    \langle t_a(0)t_b(1)\rangle=\frac{h_*^2}{4\pi^2}\left(1-\frac{\varepsilon}{2} +\mathrm{O}(\varepsilon^2)\right),\qquad
    h^2_*=N+8+\frac{4N^2+45N+170}{2(N+8)}\lsp\varepsilon+\text{O}(\varepsilon^2)\,.
\end{equation}
The critical coupling $h_*$ was computed to this order in~\cite{Allais:2014fqa, Cuomo:2021kfm} and the $1-\varepsilon/2$ comes from the renormalisation of $t$ in the MS scheme~\cite{Gimenez-Grau:2022czc} and the expansion of the free propagator at $d=4-\varepsilon$.

For $J=1$, the four-point function of tilts is in \eqref{Gtttt} and for larger $J$, the correlator of two tilts and two $\cT^J$ operators is in the ancillary \emph{Mathematica} file. This latter calculation is actually irrelevant for the integral identity \eqref{integral-id} since it requires the connected correlator and the first connected Feynman graph is at order $\varepsilon^J$ in perturbation theory. It would seem then that up to that order, the left hand side of the equation vanishes, while the right-hand side is clearly non-zero. The identity still holds once one accounts for contact terms in the four-point function~\cite{4-pot} (a somewhat different justification for the terms not captured by the smooth part of the four-point function is offered in~\cite{Cavaglia:2022qpg}).

In order to examine the contact terms, it is better to write the 
identity in terms of a double integral with an arbitrary function 
$\varphi$ satisfying $\varphi(\tau_3)=0$, $\varphi(\tau_4)=1$~\cite{4-pot}
\beq
\label{integral-id2}
\int d\tau_1\,d\tau_2\,(\varphi(\tau_1)-\varphi(\tau_2)) 
\langle t_a (\tau_1) t_b(\tau_2) 
\cT^J(\tau_3,\theta) \cT^J (\tau_4,\theta') \rangle_c
=(\theta_a\partial_{\theta^b}-\theta_b\partial_{\theta^a}) 
\langle \cT^J(\tau_3,\theta) \cT^J(\tau_4,\theta') \rangle\,.
\eeq
The contact terms picked up by the integration in \eqref{integral-id2} 
(for $\tau_3\neq\tau_4$) are proportional to 
$\delta(\tau_{14})\delta(\tau_{23})$ and $\delta(\tau_{13})\delta(\tau_{24})$. 
Less singular terms do not contribute and $\delta(\tau_{12})$ 
terms cancel due to antisymmetrisation in \eqref{integral-id2}.

To determine the contact terms, we examine the OPE of $t_a$ and $\cT^{J}$, 
which is very restrictive and takes the schematic form
\bal
\label{tTOPE}
t_a (0) \cT^{J}(\tau,\theta) &=
\frac{\lambda_{t \cT^J \cT^{J-1}}}{|\tau|^{2+\varepsilon(\gamma_{\cT^J} -\gamma_{\cT^{J-1}})}} \theta_a\cT^{J-1}(0,\theta)
+\frac{\lambda_{t \cT^J \cV^J}}{|\tau|^{1+\varepsilon(\gamma_{\cT^J} -\gamma_{\cV^J})}} \theta_a\cV^J(0,\theta) 
\\&\quad
+\frac{\lambda_{t \cT^J \cT^{J+1}}}{(J+1)|\tau|^{\gamma_{\cT^J} -\gamma_{\cT^{J+1}}}} \partial_a\cT^{J+1}(0,\theta) +\cdots.
\eal
The prefactor of $\cT^{J-1}$ has the form
\begin{equation}
\frac{\varepsilon}{|\tau|^{2+\varepsilon\gamma}}\,,
\end{equation}
which has a double pole at $\tau=0$, but no distributional part.%
\footnote{We thank P. Kravchuk for clarifying this point.}
For the second term
\begin{equation}
\frac{\varepsilon}{|\tau|^{1+\varepsilon\gamma}}
\sim\frac{2}{\gamma}\delta(\tau)+\frac{\varepsilon}{|\tau|}+\cdots,
\end{equation}
and the $\cT^{J+1}$ term is regular. So only the $\cV^J$ term 
contributes to the contact terms and we find
\bal
\label{4ptcontact}
\langle t_a (\tau_1) t_b(\tau_2) 
&\cT^J(\tau_3,\theta) \cT^J (\tau_4,\theta') \rangle
\Big|_{c,\varepsilon^0}
\\&\simeq
\left(\frac{2\lambda_{t \cT^J \cV^J}}
{\varepsilon(\gamma_{\cT^J} -\gamma_{\cV^J})}\right)^2
(\delta(\tau_{14})\delta(\tau_{23}) \theta_a \theta'_b +\delta(\tau_{13})\delta(\tau_{24}) \theta_b \theta'_a)\langle \cV^J(\tau_3,\theta) \cV^J(\tau_4,\theta') \rangle\,.
\eal
In this expression we omitted possible $\delta(\tau_{12})$ terms that drop out by the symmetrisation in \eqref{integral-id2}.

Plugging \eqref{4ptcontact} into \eqref{integral-id2} we find
\begin{equation}
\label{firstintresult}
\frac{4\lambda_{t\cT^J \cV^J}^2}{\varepsilon^2(\gamma_{\cT^J}- \gamma_{\cV^J})^2}=J\,.
\end{equation}
Given the anomalous dimensions in Table~\ref{tab:dimJ}, the structure 
constant is
\begin{equation}
\label{largeJlambda}
\lambda_{t\cT^J \cV^J}
=\frac{\sqrt{J}}{2}\,\varepsilon\,.
\end{equation}
This is confirmed by direct Feynman diagram calculation.

The analysis can be repeated for the four-point function of two tilts and two 
$\cV^J$ operators. Now the requisite OPE is
\bal
\label{tVope}
t_a (\tau_1) \cV^J (\tau_2,\theta) &=
\frac{\lambda_{t \cV^J \cV^{J-2}}}
{|\tau_{12}|^{2+\varepsilon(\gamma_{\cV^J} -\gamma_{\cV^{J-2}})}}
\theta_a \cV^{J-2}(\tau_1,\theta) 
+\sum_M\frac{\lambda_{t \cV^J\sigma_M^{J}}}
{|\tau_{12}|^{1+\varepsilon(\gamma_{\cV^J} -\gamma_{\sigma_M^{J}})}}
\theta_a \sigma_M^{J}(\tau_1,\theta)
\\&\quad
+ \frac{\lambda_{t \cV^J \cT^{J}}}
{J|\tau_{12}|^{1+\varepsilon(\gamma_{\cV^J} -\gamma_{\cT^{J}})}}
\partial_{\theta^a} \cT^{J}(\tau_1,\theta) +\cdots.
\eal
The same integral identity \eqref{integral-id2} holds, 
where the $\theta$ derivative on the right-hand side now gives a factor 
of $J-1$. At order $\varepsilon^0$ the only contribution is from the 
contact terms and amounts to
\bal
-\frac{4 \lambda_{t\cT^J \cV^J}^2}{\varepsilon^2 (\gamma_{\cT^J} -\gamma_{\cV^J})^2} \frac{J-1}{J} \left(1+\frac{2}{N+2J-5}\right) 
+\sum_M\frac{4 \lambda_{t \cV^J \sigma^J_M}^2}{\varepsilon^2 (\gamma_{\sigma^J_M} -\gamma_{\cV^J})^2}
=J-1\,,
\eal
and plugging in the expression for $\lambda_{t\cT^J \cV^J}$ in \eqref{largeJlambda}, one can rewrite it as
\bal
\label{lambda-from-int}
\sum_M\frac{\lambda_{t \cV^J \sigma^J_M}^2}{\varepsilon^2 (\gamma_{\sigma^J_M} -\gamma_{\cV^J})^2}
=\frac{J-1}{2} \left(1+\frac{1}{N+2J-5}\right).
\eal
Note that these structure constants are of order $\varepsilon$ and as 
explained at the end of Section~\ref{sec:analytic}, are not fully 
determined from a simple application of the bootstrap. For $J=2$ we 
have $\sigma^2_M\to s_\pm$ and using Feynman diagrams we evaluated
\beq
\lambda_{tVs_\pm}= 
\frac{N+20 \pm\sqrt{N^2+40N+320}}{2\sqrt{N^2+40N+320\pm (N+18)\sqrt{N^2+40N+320}}}\lsp\varepsilon\,.
\eeq
This expression is consistent with \eqref{lambda-from-int}.

For general $J$, equation \eqref{lambda-from-int} is 
yet another averaged relation for structure constants and anomalous 
dimensions. These equations complement the incomplete set of equations 
from the bootstrap for the structure constants at order $\varepsilon$ 
and so could be combined into the unmixing algorithm in 
Section~\ref{sec:analytic}.

The calculation above gives relations among the anomalous dimensions and structure constants. As already mentioned, the connected correlator at separated points vanishes up to $\text{O}(\varepsilon^J)$, so relations like \eqref{firstintresult} for $\lambda_{t\cT^J \cV^J}$ and $\gamma_{\cT^J}$, $\gamma_{\cV^J}$ to that order. For $J=1$, the first connected graph is at order $\varepsilon$, but after anti-symmetrisation the smooth part still vanishes. Given that $\Delta_\phi$ is known to order $\varepsilon^2$, this determines $\lambda_{tt\phi}$ to the same order~\cite{Sakkas:2024dvm}.

\section{Conclusion}
\label{sec:conclusion}
The magnetic line defect of the $O(N)$ model in $d=3$ is an important physical observable, but is also valuable in $d=4-\varepsilon$ as a laboratory to develop defect CFT techniques perturbatively in $\varepsilon$. In this work we presented a very detailed analysis of its low lying spectrum as well as large charge sector. This relied on Feynman diagram calculations of four-point functions of operators up to dimension two and an efficient algorithm to unmix and extract the operators exchanged in their OPEs.

All the results derived are perturbative in $\varepsilon$ and rely on Feynman diagrams. It is worth comparing the method used here to direct evaluation of anomalous dimensions and structure constants of all operators presented in Section~\ref{sec:spectrum} via renormalising and diagonalising two-point and three-point functions. To implement our algorithm (with extensive redundancies), we computed around seventy four-point functions. This is a large number, but the computations are extremely simple and in fact algorithmic (though we did not fully automate them).

The most subtle ingredient required in the calculation is the renormalisation factors of the external operators, of which we had only eight (see Tables~\ref{tab:dim1}, \ref{tab:dim2}), and only in one case (of $s_+$, $s_-$) was operator mixing important. So only eleven two-point functions had to be calculated. Additionally, as mentioned in Section~\ref{sec:spectrum}, two operators of dimension close to three do not appear in any OPE channel of lower dimensional operators and also their dimensions had to be calculated directly along with their renormalisation factors. For comparison, to compute the mixing matrices of the seven nearly degenerate scalars $s''_M$ in the first row of Table~\ref{tab:dim4} would require 28 two-point functions.

The computational complexity does not seem significantly different, with our method offering the advantage of multiple independent consistency checks and providing some information about structure constants without extra effort. The clear disadvantage is that it provides the spectrum, but not the map between operators and their expression in terms of basic fields. This still demands the evaluation of the renormalisation factors and so requires them to be operators in a Feynman diagram calculation. Higher-point functions could substitute the need to perform such explicit renormalisations of new operators. For example, six-point functions of dimension-one operators may help unmix dimension-three operators, similarly to how we were able to unmix dimension-two operators using four-point functions of dimension-one operators.

This method can easily be applied to other defects in the same model \cite{Trepanier:2023tvb, Raviv-Moshe:2023yvq, Giombi:2023dqs, Harribey:2023xyv, Harribey:2024gjn, deSabbata:2024xwn, Anataichuk:2025zoq} or to other line defects in different models \cite{Pannell:2023pwz}. Likewise, it can be applied to the large $N$ expansion of this and other models. In another context, it may prove useful in extracting data in long-range models as discussed in recent literature \cite{Benedetti:2024wgx}.

The spectrum presented here could have further applications in the study of this defect. The inversion formula for line defects was developed in~\cite{Mazac:2018qmi}, but the explicit form of the inversion kernel is known only in very limited examples. As the spectrum should arise from the inversion formula, our results should constrain the kernels and we hope that they could serve as a guide to finding them.

Similarly, four-point functions at order $\varepsilon^2$ can be engineered from particular information on exchanged operators at order $\varepsilon$. This detailed study of the spectrum could also feed into such calculations and provide more information about this defect.

\ack{We are grateful to P.~Kravchuk and J.~Mann for useful discussions, and to G.~Sakkas and R.~Rodrigues for collaboration in different stages of this project. ND acknowledges the hospitality of CERN, DESY, EPFL, Perimeter Institute and the Simons Center for Geometry and Physics in the course of this work. ND's research is supported in part by the Science Technology \& Facilities council under the grants ST/P000258/1 and ST/X000753/1. ZK is supported by ERC-2021-CoG-BrokenSymmetries 101044226. AS thanks the CERN Department of Theoretical Physics for hospitality in the course of this work. AS is supported by the Royal Society under grant URF\textbackslash{}R1\textbackslash211417 and by STFC under grant ST/X000753/1.}

\bibliography{refs}

\end{document}